\documentclass[twocolumn, showpacs, amsmath, amssymb, prb]{revtex4}

\usepackage{amssymb}
\usepackage{amsmath}
\usepackage{amsfonts}
\usepackage{graphicx}

\makeatother

\newcommand{\be}{\begin{equation}}
\newcommand{\ee}{\end{equation}}
\newcommand{\ba}{\begin{eqnarray}}
\newcommand{\ea}{\end{eqnarray}}

\begin{document}

\title{Radiative damping of surface plasmon resonance in spheroidal
metallic nanoparticle embedded in a dielectric medium}

\author{Nicolas I.~Grigorchuk \footnote{email: ngrigor@bitp.kiev.ua}}

\affiliation{Bogolyubov Institute for Theoretical Physics,
National Academy of Sciences of Ukraine, \\
14-b Metrologichna Str., Kyiv-143, Ukraine, 03680}

\begin{abstract}
The local field approach and kinetic equation method is applied to
calculate the surface plasmon radiative damping in a spheroidal metal
nanoparticle embedded in any dielectric media. The radiative damping of
the surface plasmon resonance as a function of the particle radius, shape,
dielectric constant of the surrounding medium and the light frequency is
studied in detail. It is found that the radiative damping grows quadratically
with the particle radius and oscillates with altering both the particle
size and the dielectric constant of a surrounding medium. Much attention
is paid to the electron surface-scattering contribution to the plasmon
decay. All calculations of the radiative damping are illustrated
by examples on the Au and Na nanoparticles.
\end{abstract}

\pacs{78.67.Bf; 68.49.Jk; 73.23.-b; 78.67.-n; 52.25.Os; 36.40.Vz; 25.20.Dc; 52.25.Os; 78.47.+p.}

\date{\today}

\maketitle


\newpage
\section{Introduction}
\label{intro}

The surface plasmons (SPs) excited by an electromagnetic radiation falling
on a metal nanoparticle (MN) are still of great fundamental interest\cite{SRJ,MP,LSK}
since their pronounced local resonances, whose position, shape and intensity can
be tuned over wide spectral range by varying the size and shape of the MNs or
by changing the surrounding medium.

Once excited, plasmon oscillations can damp non-radiatively by absorption
caused by electron-phonon interactions, and/or radiatively by the resonant
scattering process.\cite{BH,KV} The electron-phonon interaction with decreasing MN size becomes
more and more ineffective due to kinematical restrictions imposed on the energy-momentum
conservation laws.\cite{BGT} Nevertheless, the line broadening has been observed
in experiments with MNs scattering and absorption.\cite{KHM,NGP,LSK,SFW,MBS,DSP,SPO,Lak,HNF}
This means that besides electron-phonon interaction the other damping mechanisms should
be studied as well in order to interpret the observed SP line broadening.

Usually, both the surface and the radiative damping mechanisms play an important
role in the SP decay. In Ref.~\onlinecite{HNF} it was shown that both the electron
surface scattering and radiative damping can make significant contribution to
the linewidth $\Gamma$ (the full width at a half of maximum of the surface
plasmon resonance (SPR)).

In the MNs of a smaller radii, the penetration depth of the plasmon field reduces and
becomes more localized near the surface.\cite{KV} This is due to the fact that the electron
screening is increased as the particle radius is reduced. As a result, the bulk-induced
loss processes play only a minor role and the electronic excitations generated by the
surface potential dominate. On the other hand, if electrons are confined in
nanoparticles, where the mean free path (MFP) of the electrons becomes comparable
with the size of the particles, the electron-surface scattering comes
into play\cite{KV} and the surface acts as an additional scatterer.

The radiative damping of the SPR is an important parameter since it would
help to analyze the specificity of the transformation of the collective
electron oscillation energy into the optical far field.

The interest to treat of the SPR and radiative damping is maintained
as well by the investigation of the local field enhancement,
an effect which increases the intensity of the incident light
near the MNs surface by several orders of magnitude.\cite{BH,KV}
Since a lot of devices incorporating MNs gains from this effect, $\Gamma$
is treated as a main parameter in applications such as field concentration for
nanopatterning with nanowires,\cite{LX} plasmonic nanolithography,\cite{SFS}
and near-field optical microscopy,\cite{SD} astigmatic optical tweezers,\cite{APS}
surface enhanced Raman scattering\cite{Nie} etc.

The damping of SPR in MNs previously has been well studied
theoretically.\cite{SH,WGL,MW,MWJ,CS,KCZ,LPG}
The influence of a nearby surface on the plasmon resonance of a metallic nanoparticle of finite size
has been studied in the work \onlinecite{SH} with an account for the effects of both radiative
and evanescent surface-reflected waves.
In Ref. \onlinecite{MWJ} the oscillations of the plasmon linewidth as a function of the radius of the nanoparticles were obtained
from numerical calculations based on the time dependent local density approximation without regard for radiative damping.
The temperature effect on the radiative lifetime of the surface plasmon was studied in the work \onlinecite{LPG}.
To evaluate size and shape dependent dielectric functions and extinction spectra for MNs
with various shape, the different empirical formulas were proposed for electron MFP.\cite{CS}

The radiative damping calculated in Refs. \onlinecite{WGL,MW,KCZ} is proportional to the product
of the polarizability (proportional to particle volume) times $k^3=(2\pi/\lambda)^3$,
where $k$ is wave number and $\lambda$ is the wavelength.
The calculations have been conducted in the electrostatic approximation,
assuming $V/\lambda^3 \ll 1$, where $V$ is the MN volume.

In the works mentioned above was dropped from account the fact that the role of an external electric field
(which causes the dipole oscillations of electrons and the radiative damping) can play the inner electric
field in MN. Besides, in these works the surface and radiative decay have not been considered
simultaneously for nonspherical MN.

The purpose of the present paper is to calculate the radiative damping
of the surface plasmon under the effect of inner electric field, for the 
case when the MFP of the electrons is larger than the particle size and 
their scattering on the particle surface plays an important role.

Within the framework of a local
field approach and the kinetic equation method, we have found that the
linewidth of SPR increases quadratically with the particle radius and
exhibits oscillations as a function of the nanoparticle radius
and the refractive index of a surrounding medium.

The rest of the paper is organized as follows.
In Section II the theoretical background to the problem is presented.
Section III is devoted to the study of radiative damping. Section IV contains
the calculation of the conductivity tensor and the discussion of obtained
results. Section V contains the summary and conclusions.

\section{Surface Plasmon Linewidth}

An interaction of light with a MN embedded in a medium is studied
in the framework of classical optics, assuming that the particle and
the medium are continuous, homogeneous, and characterized by their dielectric function.
To overcome the problem connected with an inhomogeneous line broadening (due to the
size and shape distribution within the particle ensemble), we restrict ourselves
only to a {\em single} MN, which directly yields the homogeneous linewidth.\cite{KPG,SFW}

In general, the linewidth can be decomposed into contributions from the
bulk dielectric constant, surface scattering, and radiative damping:\cite{BH}
$\Gamma=\Gamma_{\rm b}+\Gamma_{\rm s}+\Gamma_{\rm rad}$.

In the classical case of free electrons in bulk metal, the damping $\Gamma_b (\equiv\nu)$
is due to the inelastic scattering of the electrons on phonons, lattice defects, or
impurities ($\nu$ refers to the electron collision frequency), which shorten the MFP.\cite{AM}
In this case, the relation $\Gamma_b = \upsilon_F/l_{\infty}$ holds, where $\upsilon_F$
is the Fermi velocity and $l_{\infty}$ is the MFP of conduction electrons in the bulk
(when the MN radius tends to infinity).

Similarly, to estimate the surface effect in electron-surface
scattering, the following empirical relation\cite{KV,CS}
\be
 \label{eq 1}
  \Gamma_s = A\frac{\upsilon_F}{l_{\rm eff}}
   \ee
is often used, where $A$ is a phenomenological factor and $l_{\rm eff}$
is a reduced effective MFP. For the sphere, the classical theory gives
$l_{\rm eff}=R$ for an isotropic scattering or $l_{\rm eff} = 4R/3$
for a diffusive scattering, while the quantum theory in a box model
yields $l_{\rm eff} = 1.16 R$ or $l_{\rm eff} = 1.33 R$,
where $R$ is the radius of a spherical MN.\cite{CS}

The bulk scattering does not cause appreciable attenuation for the high frequency
case, when $\omega\tau\gg 1$, where $\tau$ is the relaxation time. On the other hand, the contribution
to damping from surface scattering is negligible only if $l_{\infty}/2R \ll \sqrt{1+\omega^2\tau^2}$.
Thus, the surface contribution is important if $\omega\tau\gg{\rm max}(1,l_{\infty}/2R)$.

The radiative contribution can be estimated with the help of the following expression\cite{SFW}
\be
 \label{eq srad}
  \Gamma_{\rm rad} = 2\hbar k V,
   \ee
where $V$ is the nanoparticle volume and $k$ is an another
phenomenological constant to be taken from the experimental data.

To account both the surface and the radiative effects for a MN embedded
in a homogeneous, transparent medium, the dielectric permittivity\cite{HNF,CS}
\be
 \label{eq eps}
  \begin{array}{ll}&\epsilon'(\omega)=
   \epsilon_{\rm inter}(\omega)+\left(1-\frac{\omega^2_{\rm pl}}
    {\omega^2+(\Gamma_b+\Gamma_s+\Gamma_{\rm rad})^2}\right),
     \\ \\&\epsilon''(\omega) = \frac{\omega^2_{\rm pl}}{\omega} \frac{\Gamma_b+
      \Gamma_s+\Gamma_{\rm rad}}{\omega^2+(\Gamma_b+\Gamma_s+\Gamma_{\rm rad})^2}.
       \end{array}
        \ee
is usually used, where $\epsilon'$ and $\epsilon''$ are, respectively,
the real and imaginary parts of the dielectric function of the particle
material, $\epsilon_{\rm inter}$ accounts for the {\em inter}band electron
transitions, $\omega^2_{\rm pl} = 4\pi n_e e^2/m$, and $n_e$ refers
to the electron concentration. The expression in the parentheses is applied
for the {\em intra}band electron transitions.\cite{FN1}

One can see that the effect of the surface is reduced simply to the addition of a term
$\gamma_s \equiv\gamma_s(l_{\rm eff})$ in the denominator of Eq.~(\ref{eq eps}) in the form of Eq.~(\ref{eq 1}),
and the radiative effect is accounted simply by the term $\gamma_r$ in the same denominator.
Formulae (\ref{eq 1}) and (\ref{eq srad}) can be applied only to the MNs of a spherical
shape in the case when the MFP of electrons $l$ is smaller than the particle size $d$.
If the shape of MN differs from the spherical one or the inequality $l > 2R$ takes place,
then the expressions (\ref{eq 1}) and (\ref{eq srad}) can no longer be used.

 We will consider the MNs with a moderate sizes for whose, on the one hand, the condition
 $l > 2R$ is still fulfilled (the collisions of the conduction electrons with the particle
 surface remains to be an important relaxation process), and on the other hand, with the sizes
 enough large to account for the dissipation of the electron energy due to the
 emission of electromagnetic waves by plasmons, so called radiation damping.

\section{Radiative damping}

The problem of a damping of the electron energy due to the radiation
of a portion of the collective electron oscillation energy into the optical
far field has been extensively studied in the literature.\cite{KV,SFW,MW,DSP,LPG}

As we can see from Eq.~(\ref{eq 1}) the surface damping depends on a particle size.
The relative contributions from the radiative damping through the resonant
scattering and absorption also strongly depend on the particle size.
In particular, it is known\cite{BH,KV} that the plasmon absorption is
the only process taken place in small particles, whereas both the absorption
and the scattering are present in large particles, with the latter becoming
more dominant as the particle size increases. The phenomenon is based
on an interplay of the dissipative and radiative damping.

 An another assumption that the wavelength of the absorbing light
 $\lambda$ is far above the characteristic size of the nanoparticle
 (about 25 nm for gold particles\cite{KV}) allows us to treat the
 MN as being immersed in a spatial uniform, but oscillating in time
 electric field. This implies that EM field around the MN can be
 considered as homogeneous and across the particle as uniform, such
 that all the conduction electrons move in-phase producing only
 dipole-type oscillations.

The external electric field ${\bf E}^{(0)}\exp{(-i\omega t)}$
induces an inner (potential) electric field ${\bf E}_{\rm in}$ inside
the particle which is coordinate independent. The field ${\bf E}_{\rm in}$
can be linearly expressed in terms of ${\bf E}^{(0)}$ by employing
the depolarization tensor. In terms of principal axes, the depolarization
tensor, which coincide with with the principal axes of the ellipsoid,
the relation between components of external and inner fields look as\cite{BH}
\be
 \label{eq e}
  E_j^{(0)} = E_{j,{\rm in}} [1+L_j(\epsilon/\epsilon_m -1)],
   \ee
$E_{j,in}$ are the components of the electric field inside the MN and $L_j$
are the principal value of the $j$-th component of the depolarization tensor
that is also known as a geometric factors. The explicit expressions of $L_j$
for a MN with a particular shape can be found elesewhere (see, e.g.,
Refs.~[\onlinecite{Osb}], [\onlinecite{BH}],  and [\onlinecite{LL}]).
The complex dielectric permittivity of the particle material is denoted by
$\epsilon(=\epsilon'+\epsilon'')$ and $\epsilon_m$ refers to the dielectric
constant of the surrounding medium.

Electrons are accelerated in the presence of an electric field inside the MN.
It is well known from the classical electrodynamics\cite{J} that accelerated
charges emit electromagnetic radiation in all directions.
To calculate the line broadening that is entirely caused by an increase of
$\Gamma_{\rm rad}$ due to the radiative effect, we will use the time
dependence of a classical dipole oscillator. The force of a decelerative
radiation of a dipole under an inner electric field (Eq.~(\ref{eq e}))
can be represented as
\be
 \label{eq f}
  {\bf F}_{\rm rad}(t) = - \frac{2e}{3c^3}
    \sqrt{\epsilon_m}\,\left[1+L(\epsilon/\epsilon_m-1)\right]\, {\bf \dddot{d}}(t),
     \ee
where ${\bf d}$ is the MN dipole moment.
The minus sign means that this force is opposit to the dipole moment direction.
In the case of one electron and a medium with $\varepsilon_m=\varepsilon=1$,
Eq.~(\ref{eq f}) transforms into the well-known expressions from the classical
electrodynamics.\cite{J} The linewidth due to the radiative damping
of dipole vibrations can be expressed through the ${\bf F}$ by means of
\be
 \label{eq gf}
  \Gamma_{\rm rad} = \frac{e}{m}{\rm Im}
   \left[\frac{{\bf F}_{\rm rad}(t)}{{\bf \dot{d}}(t)}\right] N.
    \ee
where $N = V n_e$ is the number of free electrons
in the MN and $V$ refers to the particle volume. It is necessary to underline
here that the radiation damping rate is proportional to the total number of
oscillating electrons in MN.

Supposing ${\bf d}(t)={\bf d}_0\exp{(-i\omega t)}$, we obtain for
$j$-th component of a radiative linewidth the following expression:
\be
 \label{eq gam}
  \Gamma_{j,\rm rad} = \frac{2}{3}\frac{e^2\omega^2}{mc^3}N
   \sqrt{\epsilon_m} \,\,{\rm Im}\left[1+L_j(\epsilon_{jj}/\epsilon_m-1)\right].
    \ee
Putting here $\epsilon''_{jj}\equiv\epsilon_m$,
we get known expression, e.g., from Ref.~\onlinecite{LPG}.

For the sake of simplicity, we will assume that the dielectric
matrix has no influence on the MN and can be characterized by
\be
 \label{eq epsl}
  \epsilon'_m(\omega)=const\equiv\epsilon'_m, \qquad   \epsilon''_m(\omega)=0,
   \ee
i.e., the dielectric constant of the surrounding medium is assumed to be
frequency independent. However, it may happen in some actual cases that the
dielectric medium is strongly absorptive at frequencies below the $\omega_{pl}$.
If that is the case, then the $\epsilon''_m$ is strongly dependent
on frequency and contributes to the attenuation of the oscillations.
With accounting for the dielectric matrix properties given
by Eq.~(\ref{eq epsl}), Eq.~(\ref{eq gam}) can be rewritten as
\be
 \label{eq gep}
  \Gamma_{j,\rm rad}(\omega) = \frac{2}{3}\frac{e^2 \omega^2}{mc^3}N L_j
   \frac{\epsilon''_{jj}(\omega)}{\sqrt{\epsilon_m}}.
    \ee

If one will consider the case of the frequency close to the frequency
of the bulk plasma oscillations of electrons in metal $\omega_{\rm pl}$,
then the imaginary part of the dielectric function tensor for free electron
gas can be expressed within the Drude-Sommerfeld model as\cite{BH,KV,AM}
\be
 \label{eq esh}
   \epsilon''_{jj}(\omega)=4\pi\frac{\sigma'_{jj}(\omega)}{\omega},
    \ee
where  $\sigma'_{jj}$ is the principal components of the real part of the
conductivity tensor. Taking into account the expression (\ref{eq esh}),
we get
\be
 \label{eq gsi}
  \Gamma_{j,\rm rad}(\omega)=\frac{8\pi}{3}\frac{e^2 \omega}{m c^3}N L_j
   \frac{\sigma'_{jj}(\omega)}{\sqrt{\epsilon_m}}.
    \ee

The real part of the conductivity tensor can be expressed through the imaginary part 
of the polarizability tensor $\alpha_{jj}$ by means of
\be
 \label{eq sial}
  \sigma'_{jj}(\omega) = \frac{\omega}{V}\epsilon_m |1+L_j(\epsilon/
   \epsilon_m-1)|^2 \,{\rm Im}\,\alpha_{jj}(\omega).
    \ee

In the case of the SPR, one can use in Eq.~(\ref{eq eps})
for nonspherical MNs, the frequency
\be
 \label{eq prf}
  \omega=\omega_{\rm sp}=
   \frac{\omega_{\rm pl}}{\sqrt{\varepsilon_{\infty}+(1/L_j-1)n^2}}.
    \ee
Here $n^2=\epsilon_m$ and $\varepsilon_{\infty}\equiv 1+\epsilon_{\rm inter}$
is the high frequency dielectric constant due to interband and core
transitions of the inner electrons in a MN's material.

There are different possibilities to calculate $\sigma'_{jj}(\omega)$
for the different frequency regime. Below, we will demonstrate how to
calculate the $\sigma'_{jj}$ as applied to the nanoparticle case.

\section{Calculation of $\sigma'_{jj}$}

To calculate the tensor $\sigma'_{jj}$ we will use the kinetic equations method.
Benefit of this method is that it permits one to study the effect of the
particle shape on the measured physical values.
Second, it enables us to investigate the particles whose sizes are those that
the particle surface start to play an important role. A diffuse boundary
scattering is assumed to be a good approximation in this case.

The simplest form of a nonspherical shape is a spheroid.
We restrict ourselves to the nanoparticles with a {\em spheroidal shape} only.
Applying the mentioned method, we have found\cite{GT1} that the components
of the conductivity tensor for light polarized along ($\|$) or
across ($\bot$) the rotation axis of a spheroidal MN are
\be
 \label{eq sig}
  \sigma'_{\|\choose\bot}(\omega)=\frac{9\omega^2_{\rm pl}}{16\pi}{\rm Re}
   \left[\frac{1}{\nu-i\omega}\int\limits_0^{\pi/2}
    {\sin\theta\,\cos^2\theta \choose \frac{1}{2}\sin^3\theta}\Psi(\theta)
     \;d\theta\right]_{\upsilon=\upsilon_F},
      \ee
where $\nu$ is the electron collision frequency and $\theta$
is the angle between rotation axes of the spheroid and direction
of an electron velocity. Here and below, the upper (lower) symbol
in the parentheses on the left-hand side of Eq.~(\ref{eq sig})
corresponds to the upper (lower) expression in the parentheses
on the right-hand side of this equation. The subscript
$\upsilon=\upsilon_F$ means that the electron velocity in
the final expressions should be taken on the Fermi surface.

The complex $\Psi$ function entering in Eq.~(\ref{eq sig}) has the form
\be
 \label{eq psi}
  \Psi(q)=\Phi(q)-\frac{4}{q^2}\left(1+\frac{1}{q}\right)e^{-q},
   \ee
where
\be
 \label{eq phi}
  \Phi(q)=\frac{4}{3}-\frac{2}{q}+\frac{4}{q^3},
   \qquad   q=\frac{2R}{\upsilon'}(\nu-i\omega).
    \ee
One can see from Eq.~(\ref{eq phi}) that the $q$ is governed by the "deformed"
electron velocity, which in the case of a spheroidal MN takes the form
\be
 \label{eq ve}
  \upsilon'=\upsilon R\sqrt{\left(\frac{\sin\theta}{R_{\bot}}\right)^2+
   \left(\frac{\cos\theta}{R_{\|}}\right)^2}\equiv\upsilon'(\theta),
    \ee
where $\upsilon$ refers to the electron velocity in a spherical particle
with radius $R$; $R_{\|}$ and $R_{\bot}$ are the spheroid semiaxes directed
along and across the spheroid rotation axis, respectively.\cite{FN2}
The semiaxes are connected to the radius of sphere $R$ of an equivalent
volume through the relation $R^3={R_{\|}}R^2_{\bot}$.

The last summand in Eq.~(\ref{eq psi}) represents the oscillation part
of the $\Psi$ function and the first one refers to its smooth part.
The $\Psi$ function in Eq.~(\ref{eq psi}) varies with the angle $\theta$
because the parameter $q$ becomes dependent on the angle $\theta$
for a spheroidal particle, namely
\be
 \label{eq qu}
  q(\theta)=\frac{2}{\upsilon_F}
   \frac{\nu-i\omega}{\sqrt{\frac{\cos^2\theta}{R^2_{\|}}+
    \frac{\sin^2\theta}{R^2_{\bot}}}}.
     \ee

\subsection{Conductivity tensor in high-frequency limit}

Let us introduce the frequency of electron oscillations between particle walls as
\be
 \label{eq nus}
   \nu_s=\frac{\upsilon_F}{2R}.
    \ee
Depending on sizes of MN, its shape and temperature, the variety of relations
between frequencies $\nu_s$, $\nu$ and $\omega_{\rm pl}$ can be achieved.
For example, for the Na nanoparticle with the radius of $R < 2~\AA$,
$\nu_s\simeq\omega_{\rm pl}$. On the other hand, with $R > 126~\AA$,
the electron oscillation frequency becomes $\nu_s < \nu$, where
$\nu\simeq 4.24\cdot 10^{13}$~s$^{-1}$ as is estimated for the Na at 300$^0$~K.
This leads to different expressions for $\sigma(\omega)$, which can
be used in Eq.~(\ref{eq gsi}) for calculation of the plasmon linewidth.

Below, we will consider the case when the contribution of the bulk damping to the
radiative plasmon linewidth of SPR is neglected. The components of the conductivity
tensor for a spheroidal MN in the highfrequency (HF) limit ($\omega\gg\nu_s$)
and $\nu_s\gg\nu$, can be represented as\cite{GT2}
\be
 \label{eq sgs}
  \sigma'_{\|\choose\bot}(\omega)=\frac{9}{32\pi}
   \left(\frac{\omega_{\rm pl}}{\omega}\right)^2
    \frac{\upsilon_F}{R_{\bot}}{\eta(e_p) \choose \rho(e_p)},
     \ee
where $R_{\bot} (= R x^{1/3}, x=R_{\bot}/R_{\|})$ is a spheroid semiaxis
directed across to the spheroid rotation axis, and $\eta(e_p)$ and $\rho(e_p)$
are some smooth functions\cite{TG} dependent only on the spheroid eccentricity
$e_p=\sqrt{1-x^2}$ (a prolate spheroid), or $e_p= \sqrt{x^2-1}$ (an oblate one).

One can use the following asymptotic expressions for functions $\eta(x)$ and $\rho(x)$
in the cases of both the extremely small or the large axial ratio, respectively:
\be
 \label{eq etr}
  \begin{array}{ll}
   &\eta(x)\simeq\left\{
    \begin{array}{ll}
     \pi/8+3\pi x^2/16, &\textrm{ for a prolate spheroid} \\ \\
      x/2+1/(4x), & \textrm{ for an oblate spheroid}
       \end{array}
        \right.,
\\ \\
&\rho(x)\simeq\left\{
 \begin{array}{ll}
  3\pi/16+\pi x^2/32, & \textrm{for a prolate spheroid} \\ \\
   \frac{x}{4}+\frac{-1+4\ln{2x}}{8x}, & \textrm{for an oblate spheroid}
    \end{array}
     \right..
     \end{array}
      \ee

If one consider the nanowires and nanorods, which can be reasonably approximated
as prolate spheroids, then one can put $\eta\simeq\pi/8$ and $\rho\simeq 3\pi/16$
with a sufficient degree of accuracy.
In the case of MNs with a spherical shape $\eta=\rho=2/3$.

\subsubsection{Radiative damping for spheroidal MN in HF limit}

In the highfrequency limit, when Eq.~(\ref{eq sgs}) can be applied, for two
components of the linewidth of a spheroidal MN embedded in a medium with $\epsilon_m$,
 we obtain  the following equation
\be
 \label{eq gsp}
  \Gamma_{{\|\choose\bot},{\rm rad, sp}}=\frac{3}{4}\frac{e^2\omega^2_{\rm pl}}{m\omega c^3}
   \frac{N \upsilon_F}{R_{\bot}\sqrt{\epsilon_m}} L_{\|\choose\bot} {\eta(e_p)\choose\rho(e_p)},
     \ee
where explicit expressions for $\eta(e_p)$ and $\rho(e_p)$ can be found
in Ref.~\onlinecite{TG}. Accounting for
\be
 \label{eq n}
   N = \frac{4}{3}\pi R_{\|} R^2_{\bot} n_e =
    \frac{m}{3 e^2}\omega^2_{\rm pl} R_{\|} R^2_{\bot},
     \ee
Eq.~(\ref{eq gsp}) at the resonance frequency (\ref{eq prf}) takes the form
\be
 \label{eq gan}
  \Gamma_{{\|\choose\bot},{\rm rad, sp}} = \frac{\omega^3_{\rm pl}}{4 c^3}
   \upsilon_F R_{\|} R_{\bot} L_{\|\choose\bot}
    \sqrt{\frac{2n^2+\epsilon_{\infty}}{n^2}} {\eta(e_p)\choose\rho(e_p)}.
     \ee
The depolarization coefficients $L_{\|\choose\bot}$ for spheroidal MN can be found elsewhere.\cite{BH,KV}
In the case of MNs with a very elongated shape ($x\ll 1$), they can be expressed as
\be
 \label{eq dep}
  L_{\|}(x)\simeq x^2 \left[\ln{\left(\frac{2}{x}-\frac{x}{4}\right)}-1\right],
   \quad  L_{\bot}(x)=[1-L_{\|}(x)]/2.
    \ee

\subsubsection{Radiative damping for a spherical MN in HF limit}

In the case of MN with a spherical shape $R_{\|}=R_{\bot}\equiv R$,
and one can put the depolarization factor equal to $L_{\|}=L_{\bot}=1/3$
in Eqs.~(\ref{eq e}), (\ref{eq f}), (\ref{eq gam})--(\ref{eq prf}),
(\ref{eq gsi}), and (\ref{eq gsp}).

Eq.~(\ref{eq gan}) for MN with a spherical shape can be rewritten as
\be
 \label{eq grf}
  \Gamma_{\rm rad, sp}=\frac{1}{18}\left(\frac{\omega_{\rm pl}}{c}\right)^3
   \upsilon_F \,R^2 \sqrt{\frac{{2n^2+\varepsilon_\infty}}{n^2}}.
    \ee

The increase in linewidth from the radiative damping (as we can see from Eqs.~(\ref{eq gan})
and (\ref{eq grf})) is proportional to the MN surface area. This can be understood
from the fact that the surface scattering is a solely mechanism for the change
of the electron velocity and acceleration. We do not account any other radiative
mechanisms in our theory as, for instance, the electron-electron collisions.

The estimations of $\Gamma$ for the spherical Au ($\epsilon_{\infty} = 9.84$)
and Na ($\epsilon_{\infty} = 1.14$)\cite{K} particles with $2R = 400\AA$
embedded in the vacuum ($n = 1$), in accordance with Eq.~(\ref{eq grf}),
give: 6.7 and 0.795~meV (or 98 and 1209~fs), respectively.

If the MN is embedded in the dielectric media with $\epsilon_m > 1$,
then the environment effect ought to be taken into account.

\subsection{Environment effect}

Because the effect of an electric field on the embedded nanoparticles
becomes weaker in a dielectric media proportionally to its refractive
index, the environment effect plays an important role.
The spectral peculiarities of an environment effect recently were investigated
for the Ag and Au nanoparticles, for instance, in Refs.~[\onlinecite{ML,KCZ}].

In order to study the significance of the radiative linewidth behavior
in more general situations as above presented, it is necessary to perform the
numerical calculation in Eq.~(\ref{eq gsi}) with the use of a general expression
for the conductivity tensor given by Eq.~(\ref{eq sig}). Then, the
radiative linewidth of SPR with an account for Eqs.~(\ref{eq prf})
and (\ref{eq n}) can be expressed in the form
\be
 \label{eq gco}
  \Gamma_{\rm rad, {\|\choose\bot}} = \frac{8\pi}{9n}
   \left(\frac{\omega_{\rm pl}}{c}\right)^3
    \frac{R_{\|} R^2_{\bot}L_{\|\choose\bot}\sigma'_{\|\choose\bot}} {\sqrt{\epsilon_{\infty}+(1/L_{\|\choose\bot}-1)n^2}}.
      \ee
It allows to account an other important factor, namely the
dependence of the radiative damping on the particle shape.
As already was outlined,\cite{BH,KV,G} any change of the nanoparticle
shape from a sphere, that introducing of an anisotropy, results
in the splitting of the SPR into two modes: a transverse one
($\Gamma_{\bot}$, perpendicular to the spheroid axis of a revolution)
and a longitudinal one ($\Gamma_{\|}$, parallel to this axis).

In Fig.~1 we illustrate the behavior of the radiative damping components
as a function of the medium refractive index for a fixed particle axes ratio.

The calculations were conducted for the Au nanoparticle with
the use of Eqs.~(\ref{eq gco}) and (\ref{eq sig}), and the following
parameters\cite{K}: $n_e\simeq 5.9\times 10^{22}$~cm$^{-3}$,
$\upsilon_F = 1.39\times 10^{8}$~cm/s,
$\omega_{\rm pl} = 1.37\times 10^{16}$~s$^{-1}$
and $\epsilon_{\infty}=9.84$.

\noindent\includegraphics[width=8.6cm]{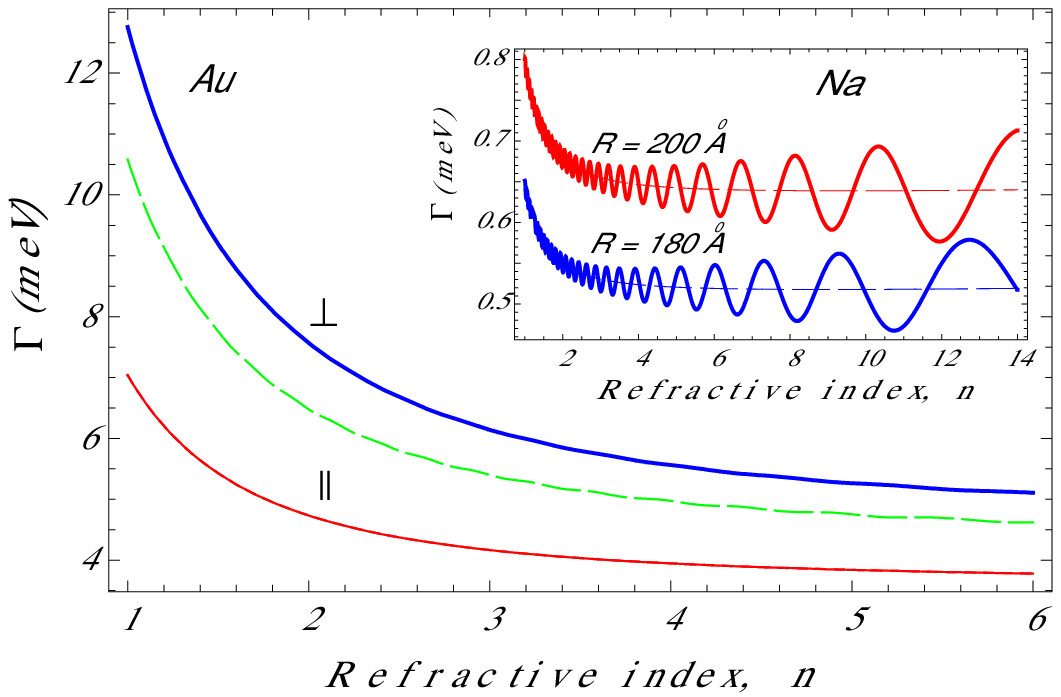}
\vskip-1mm\noindent{\footnotesize FIG.~1. (Color online)
The radiative linewidth of the surface plasmon resonance components
(longitudinal $\|$ and transverse $\bot$) vs the medium refractive
index for a prolate Au nanoparticles with the axes ratio
$R_{\bot}/R_{\|}=0.618$. The dashed line corresponds to the
spherical Au particle with the radius $R=200\AA$. The inset shows
the same dependence for two spherical Na nanoparticles with the radii
of 180 and $200 \AA$ (solid lines). The dashed lines represent the $\Gamma(n)$
in the case, when the oscillation terms in Eq.~(\ref{eq gs}) are neglected.}
\vskip10pt

We observe from Fig.~1 that both radiative damping components are
decreased with increasing the medium refractive index, in other words,
the effect is weaker for higher dielectric constant of environment.
The more of MN radius or the less the refractive index
is, the higher is the calculated radiative damping.

In general, the resonance plasmon damping in the {\em prolate}
Au nanoparticle was found to be weaker along the spheroid revolution
axis than the one across this axis. For the {\em oblate} Au
nanoparticle, on the contrary, the damping along the revolution
axis was stronger than that across this axis. This result
holds regardless of whether the photo-excitation is close
to the SPR or far from it.

The resonance peak position is shifted with changing
the refractive index of the surrounding medium.
The spectral direction of the shift depends on a number of factors,
well studied in earlier publications.\cite{BH,KV,HNF,ML,KCZ}

\subsubsection{Radiative plasmon linewidth for a spherical MN in a more general case}

We consider here the case $\upsilon'= \upsilon$,
when the $\Psi$ function ceases to depend on the angle $\theta$.
Then the integration over $\theta$ in Eq.~(\ref{eq sig})
gives 1/3 and we derive for $\sigma'$
\be
 \label{eq spc}
  \sigma'_{\rm sph}=\frac{3\omega^2_{\rm pl}}{16\pi}{\rm Re}
   \left(\frac{\Psi(q)}{\nu-i\omega}\right),
    \ee
with $q = 2R(\nu-i\omega)/\upsilon$ taken at $\upsilon = \upsilon_F$.

Let us choose the case $\nu\ll\nu_s$, for illustration.
After simple algebraic calculations with the use of Eqs.~(\ref{eq psi})
and (\ref{eq phi}), we obtain in this case:
\be
 \label{eq ssp}
  \sigma'_{\rm sph}\simeq\frac{3}{8\pi}\nu_s
   \frac{\omega^2_{\rm pl}}{\omega^2}\left[1-
    \frac{2\nu_s}{\omega}\sin\frac{\omega}{\nu_s}+
     \frac{2\nu^2_s}{\omega^2}\left(1-
      \cos\frac{\omega}{\nu_s}\right)\right].
       \ee
Substituting Eqs.~(\ref{eq ssp}) and (\ref{eq n})
into Eq.~(\ref{eq gsi}), we get
\be
 \label{eq gs}
  \Gamma_{\rm sph}\simeq\frac{\omega_{\rm pl}}{9n\xi}
   \left(\frac{R\omega_{\rm pl}}{c}\right)^3
    \left[1-\frac{2}{\xi}\sin{\xi}+\frac{2}{\xi^2}(1-\cos{\xi})\right],
     \ee
with
\be
 \label{eq xi}
  \xi\equiv\xi(n,R)=\frac{2R\omega_{\rm pl}}{\upsilon_F
   \sqrt{2n^2+\varepsilon_{\infty}}}.
    \ee

One can see from Eq.~(\ref{eq gs}) that the radiative damping
in a spherical MN oscillates with altering of both the radius
of MN and/or the refractive index of an embedding medium.
Eq.~(\ref{eq gs}) at $\xi\gg 1$ transforms into Eq.~(\ref{eq grf}).

The inset at the Fig.~1 shows the behavior of the SPR linewidth for spherical
Na nanoparticles with different radii. For calculation we use Eq.~(\ref{eq gs})
and the following parameters: $n_e\simeq 2.65\times 10^{22}$~cm$^{-3}$,
$\upsilon_F = 1.07\times 10^{8}$~cm/s,
$\omega_{\rm pl} = 9.18\times 10^{15}$~s$^{-1}$,
and $\epsilon_{\infty}=1.14$.

The dashed (smooth) curves in the inset in Fig.~1 correspond to the simple
case when both the second and the last terms are neglected in the square
brackets of Eq.~(\ref{eq gs}). In this simple case, the radiative linewidth of
the SPR is decreased with enhancing of a refractive index $n$ of the environment.
This takes place because the radiative plasmon decay is weakened in the media with
the higher refractive index owing to decrease of an external EM field inside the MN.

As one can see from the inset in Fig.~1, the oscillations of $\Gamma$ around
its smooth curve make significant corrections to the smooth picture,
especially at large $n$. The oscillations are well pronounced
for the Na nanoparticles with the small radii and are disappeared
for NP with a larger radii. This can be connected with the number of an electron
oscillations between particle walls, which is decreased as the particle radius
is increased. These oscillations is damped markedly
with $n$ decreasing and practically disappeared at $n<2$.
The damping enhances as the radius of MN becomes larger.

\subsubsection{Effect of the particle size}

Figure~2 depicts the behavior of the radiative
linewidth of SPR as the function of MN radius for some fixed
medium refractive indices. The calculations were fulfilled
for the Au nanoparticle with the use of Eq.~(\ref{eq gs})
and the same numerical parameters as given above.

 As we can see in Fig.~2 and from Eq.~(\ref{eq gs}),
 the $\Gamma$ is increased quadratically with $R$ (dashed lines).
 We suppose, this is due to the fact that the radiative damping in
 the MN with a spherical shape is proportional to the area of a sphere.
 The growth in $\Gamma$ occurs slower in the media with a higher
 refractive index. The redaction of radiating damping in the media
 with higher $n$ implies a reduced dephasing of the plasmon mode.
 Because the effect of an external electric field on the embedded
 nanoparticle becomes weaker in a dielectric media with higher
 refractive index, the resonance radiative damping tends
 to decrease within the nanoparticle.

 The $\Gamma$ oscillates
 around a smooth curves with an increase of the particle radius.
 The period of these oscillations is enhanced for the MNs embedded
 in the medium with a higher refractive index.
 The magnitude of these oscillations is the greater
 the higher refractive index is. The oscillations of $\Gamma$
 with $R$ follow the quadratic dependence as well.
 The inset at the Fig.~2 shows the same dependence for Na
 nanoparticle embedded in the media with $n=1$ and $n=9$.

On the oscillations of the SP lifetime
as a function of nanoparticle size it was pointed out for the first
time in Ref.~\onlinecite{MWJ}, where the semiclassical theory
was used to evaluate the SPR in MNs.

\noindent\includegraphics[width=8.6cm]{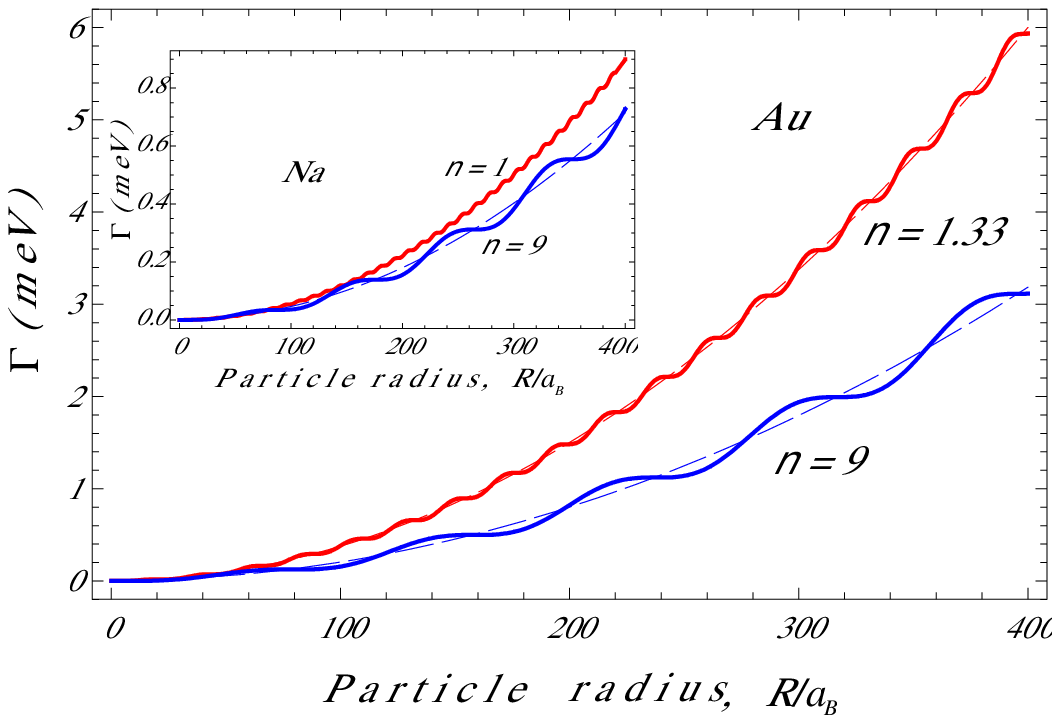}
\vskip-1mm\noindent{\footnotesize FIG.~2. (Color online)
The radiative linewidth of SPR vs radius of spherical Au
nanoparticle (in units of the Bohr radius $a_B = 0.53$)
embedded in the water $n\simeq 1.33$ or in the medium with
refractive index $n = 9$. The dashed lines represent the $\Gamma(R)$
in the case, when the oscillation terms in Eq.~(\ref{eq gs}) are neglected.
The inset shows the same dependence for Na nanoparticle.
}
\vskip10pt

The behavior of $\Gamma$ at small particle radii $R/a_B < 40$ in our
approach practically does not depend on both the particle radius
and the medium refractive index and can be described more precisely
in the frame of the quantum theory.

So far we have considered only radiative processes.
Below, we dwell shortly on the nonradiative processes when the
electron scattering in the MN dissipates oscillation energy into heat.

\subsection{Linewidth for a nonradiative processes}

Another way for calculation of $\Gamma_{\rm j, rad}$
supposes the knowledge of the value of a nonradiative
damping $\Gamma_{\rm j, nonrad}$. That is
\be
 \label{eq nonr}
  \Gamma_{\rm j, rad} = \frac{\sigma_{\rm sca}}{\sigma_{\rm abs}}
   \Gamma_{\rm j, nonrad},
    \ee
where $\sigma_{\rm sca}$ and $\sigma_{\rm abs}$ are the light
scattering and absorption cross sections, respectively.

The nonradiative damping rate of a SP can be expressed as\cite{G}
\be
 \label{eq gn}
  \Gamma_{\rm j, nonrad}(\omega)=\frac{4\pi L_{j}}
   {\epsilon'_m+L_{j}(1-\epsilon'_m)} \sigma'_{jj}(\omega),
    \ee
with the $\sigma'_{jj}$ given in the Section IV.
Eq.~(\ref{eq gn}) defines the linewidth or, correspondingly,
the decay time of the plasmon resonance due to electron scattering
both from the bulk and from surfaces of the particle.

In the case of medium with $\epsilon'_m\rightarrow 1$,
Eq.~(\ref{eq gn}) is reduced merely to
\be
 \label{eq gno}
  \Gamma_{\rm j, nonrad}(\omega)=4\pi L_{j}\;\sigma'_{jj}(\omega).
   \ee
So, the decay time of the plasmon resonance is the optical conductivity
of the MN at the light frequency multiplied by a geometrical factor.
To study the nonradiative linewidth for MNs with a given $L_{j}$, it is enough to
calculate the real part of the conductivity tensor as a function of frequency.

As one can see from Eq.~(\ref{eq nonr}), the radiative damping rate start to dominate
as the light scattering cross section by MN exceeds the light absorption one in it.
The calculations for spherical MN in vacuum gives that radius of MN for which both cross
sections become comparable each with other. In the high-frequency limit, we found that
\be
 \label{eq rHF}
  R_{\rm HF}\simeq\left(\frac{3}{2}\right)^{3/4} (c\upsilon_F)^{1/4}
   \sqrt{\frac{c}{\omega\omega_{\rm pl}}}.
    \ee
For instance, $R\simeq 100\AA$ for Au particle at the plasmon
frequency $\omega=\omega_{\rm pl}/\sqrt{3}$.

In the case of low frequencies, we have
\be
 \label{eq rLF}
  R_{\rm LF}\simeq\frac{c}{\omega_{\rm pl}}
   \sqrt{\frac{6c}{\upsilon_F\sqrt{\epsilon_m}}}.
    \ee
 The latter formula gives $R\simeq 7860\AA$ for
 Au particle embedded in a medium with $\epsilon_m=1$.

In order to take into account the radiative damping together with
collisions of free carriers with the MN surface, the effective damping rate
 $\Gamma_{\rm eff}=\Gamma_{\rm nonrad}+\Gamma_{\rm rad}$ must be introduced.
For understanding of the decay mechanism of the electron plasma oscillations
the knowledge of the decay time is of central importance.

\section{Summary and conclusions}
	
We used a local field approach and a kinetic equation method to study
the plasmon resonance linewidth for metal nonspherical nanoparticles
embedded in any dielectric media. It enables to calculate the
radiative linewidth for MNs with different geometry with an account
for the light scattering from the particle surfaces.

The general formula is proposed for a damping rate or a decay time due
to electron scattering from the bulk and particle surfaces.
By means of this formula one can evaluate the linewidth
directly through the tensor of optical conductivity of the MN.

With changing the MNs shape from spherical to the spheroidal one,
the single radiative plasmon resonance splits into two components:
the longitudinal and a transverse one to the spheroid rotation axis.
We found that both the radiative damping components for Au nanoparticle
are decreased with increasing the medium refractive index.
The resonance plasmon damping in the {\em prolate} Au nanoparticle
was found to be weaker along the spheroid revolution axis than that
across this axis.

For Na nanoparticle, we detect the oscillations of $\Gamma$ with the
refractive index increasing. The amplitude of these oscillations enhances
for media with higher dielectric constant. The oscillations are well
pronounced for Na nanoparticles with the small radii and are
disappeared for Na nanoparticles with a larger radii.

We clearly show for spherical MNs that the radiative linewidth of SPR
 enhances quadratically with the particle radius increasing.
 The growth in $\Gamma$ occurs slower in the media with a higher
 refractive index. The $\Gamma$ oscillates as well
 with changing in the particle radius.
 The magnitude of these oscillations is enhanced markedly for the MNs embedded
 in the medium with a higher refractive index.

The effects of both the particle radius and the environment on the radiative plasmon
resonance linewidth are illustrated by the example of Au and Na nanoparticles.

The contribution of the nonradiative plasmon decay is discussed as well.

Our theoretical results should be important for the analysis of the experimental
data on the optical and transport properties of MNs embedded in various dielectric media.


\begin{thebibliography}{50}
\bibitem[29]{FN1}   The latter mechanism requires a threshold energy $E_{\rm ib}$ of about 2.38~eV
                    in the Au, whereas the energy of surface plasmon in the Au is 2.29~eV.\cite{KV}
                    In the Ag $E_{\rm sp}\approx 3$~eV and $E_{\rm ib}\approx 4$~eV.\cite{KV}
\end{thebibliography}
\end{document}